\documentclass[10pt]{article}
\usepackage{citesort}
\begin{document}

\title{A coil-globule transition of a semiflexible polymer
driven by the addition of spherical particles}
\author{{\bf Richard P. Sear}\\
~\\
Department of Chemistry and Biochemistry\\
University of California, Los Angeles\\
Los Angeles, California 90024, U.S.A.\\
email: sear@chem.ucla.edu}

\date{\today}

\maketitle

\begin{abstract}
The phase behaviour of a single
large semiflexible polymer immersed in a suspension of
spherical particles is studied.
All interactions are simple excluded volume interactions and
the diameter of the spherical particles is an order of magnitude
larger than the diameter of the polymer.
The spherical particles induce a quite long ranged depletion attraction
between the segments of the polymer and this induces a continuous
coil--globule transition in the polymer.
This behaviour gives an indication of the condensing effect of
macromolecular crowding on DNA.
\end{abstract}

%\newpage
\section{Introduction}

Phase separation and partitioning driven by excluded volume interactions
have been well studied theoretically
\cite{vrij76,dijkstra94,bolhuis94,poon95,lekkerkerker95,vanroij96,searm97,bolhuis97},
with the inspiration coming from experiments both on synthetic colloidal
systems \cite{poon95,lekkerkerker95,dinsmore96} and on biologically derived
systems \cite{walter95,minton95}.
When excluded volume effects are dominant the properties of a mixture
are determined solely by the sizes and shapes of its components.
For example, mixtures of long narrow rod-like
particles and spheres have been shown to demix
solely because of these differences in size and shape \cite{bolhuis94,searm97}.
The rod-like particles could be a minimal model of a semiflexible polymer
or of a micelle, and the spheres could be small colloidal particles or
even compact protein molecules.
But if the semiflexible polymer is very long then even a single, isolated
molecule can undergo phase transitions 
\cite{degennes,lifshitz78,williams81,grosberg92,cloizeux90},
because it is then large enough,
has enough degrees of freedom, to be treated as a thermodynamic
system \cite{degennes}. Here we study such a polymer, mixed with spheres,
and see if the presence of the spheres can induce a phase transition.
We find that they can. When the concentration of spheres
exceeds a critical value the polymer molecule
contracts and expels the
spherical particles. 
In effect the polymer molecule and the spheres are demixing.
The mixture of many short rod-like polymers and spheres
demixed to form a phase with a high density of rod-like polymers but a low
density of spheres (coexisting with
one with high sphere density and low rod density)
and here the polymer contracts to form a dense
globule with a high density of polymer but a low density of spheres.
This dense phase of a polymer is referred to as the globular phase 
\cite{degennes,williams81,williams81,lifshitz78,grosberg92,cloizeux90} and 
the contraction of the polymer is the coil--globule transition.

In the work presented here, we will draw on existing theories for the demixing
of spheres and semiflexible polymers \cite{searm97} and
for the coil--globule
transition \cite{lifshitz78,grosberg92}.
The theory for the mixture of spheres and semiflexible polymers
\cite{searm97}
is a straightforward virial expansion of the Helmholtz free energy, truncated
after the second virial coefficient terms. 
In the free energy expressions of Ref. \cite{searm97} the two components
of the mixture were treated symmetrically. While this is of course
perfectly valid, here we want to use the analogy between demixing and the
vapour--liquid phase separation of a pure substance 
\cite{vrij76,poon95,lekkerkerker95,lekkerkerker92}.
To do this we will Legendre transform
\cite{chandler87} the Helmholtz free energy into a
semi-grand potential \cite{lekkerkerker92,sear95}.
This semi-grand potential is then a function of the density of the polymer
and the chemical potential of the spheres. A chemical potential, like the
temperature, is uniform throughout any system at equilibrium;
it is a field variable not a density.
So, the semi-grand potential has the same form as the Helmholtz free energy
of a polymer which interacts via interactions which are not solely
excluded volume
and so depend on temperature.
Both
depend on the density of polymer and on a field variable: temperature
in the case of a polymer with soft interactions, and the chemical potential
of the spheres for the polymer mixed with spheres.
In particular, both thermodynamic functions can be expanded as a virial
series in the density of polymer with coefficients which depend on
temperature/chemical potential of the spheres.
Just as reducing the temperature of a polymer molecule
with attractive interactions
can make some of its
virial coefficients negative, increasing the chemical potential
of spheres drives some of the virial coefficients of the semi-grand
potential of the mixture of polymer and spheres negative.
In both cases the effect is the same: the polymer contracts from the coil
state to the globular state. The spheres have in effect induced an attraction,
often called a depletion attraction, between the
segments of the polymer molecule.

The interactions between the segments of a polymer molecule determines its
state. Note that we always consider a single, isolated polymer molecule.
If the interactions between the polymer segments are repulsive, the good
solvent regime, then the polymer exists as a swollen coil \cite{degennes}.
The radius of gyration of the polymer, a measure of its size,
scales with the number of segments $N$ as $N^{3/5}$
(actually the exponent is slightly less than $3/5$ \cite{cloizeux90}).
So, the volume occupied by the polymer molecule scales as $N^{9/5}$.
The exponent is greater than one
and so the average density of segments
inside this volume tends to 0 as $N$ tends
to infinity.
However, if the interactions between the segments of the polymer are
sufficiently attractive, the poor solvent regime, the polymer exists in a
condensed state, the globular state 
\cite{degennes,lifshitz78,grosberg92,cloizeux90}.
There the radius of gyration of the
polymer scales as $N^{1/3}$ and so as $N$ tends to infinity the average
density inside the polymer remains non-zero.
The crossover from the radius of gyration scaling as $N^{3/5}$ to
$N^{1/3}$ marks the coil--globule transition.

Motivation for studying the current model mixture is provided by
an interest in the phase behaviour of long DNA double helices.
The DNA double helix is a semiflexible polymer, its
persistence length \cite{vroege92} is $\sim 50$nm \cite{merchant94} which
is 25 times its diameter of $\sim 2$nm. Our semiflexible polymer with
only excluded volume interactions is a crude model of DNA in a good solvent
\cite{vroege92,merchant94}. 
Our model mixture is then a crude but not
unreasonable model of a long DNA molecule in a suspension of spherical
particles, in the absence of any specific DNA--spherical-particle
attraction.
Thus, our results imply that DNA can be condensed using colloidal
spheres.
As far as we are aware this has not been attempted. However, there
are a number of experimental techniques for condensing
DNA, such as altering the solvent, adding polyvalent salts etc., see
Refs. \cite{ueda96,yoshikawa96} and references therein.
However, all these techniques produce a sudden collapse of the DNA
to a dense state in which the separation between adjacent parts
of the DNA is only a few nm and the DNA has hexagonal order.
The collapse induced by the colloidal spheres is continuous and
and it is then possible to prepare a low density isotropic
globule of DNA.
As far as we are aware this is the only way of preparing a low
density globule of a semiflexible polymer such as DNA.
Our system may also be useful as a very crude model of the effect
of `macromolecular crowding' on DNA actually in cells, see
Refs. \cite{walter95,minton95,murphy95} and references therein.

\section{Model}

The polymer is modeled by a
homogeneous cylindrical elastic filament \cite{vroege92}.
The filament follows a continuous curve in space, see Fig. 1. 
The filament bends and flexes during thermal motion but it
has a certain amount of rigidity, measured by its persistence length $P$
\cite{vroege92}. A piece of polymer shorter than the persistence length
only bends by a small amount due to thermal motion, it behaves almost
like a rigid rod.
The polymer has a hard core of diameter $D$, that is the centreline
of the polymer cannot approach within $D$ of itself. 
The polymer is $N$ persistence lengths long; in our calculations
we will always consider the limit of $N\rightarrow\infty$.

The colloidal spheres are modeled by hard spheres with a diameter
$D_s$.
The polymer--sphere interaction is also an excluded--volume interaction.
The centre of a sphere cannot approach within $(D+D_s)/2$ of the
centreline of the polymer.

\section{Theory}

We only consider explicitly the globular state of a single isolated
polymer molecule in the
$N\rightarrow\infty$ limit.
We also neglect any variation in the density of polymer segments in the
globule; the volume approximation of Lifshitz and coworkers
\cite{lifshitz78,grosberg92}.
Then the globule is simply a bulk phase of volume $V$ in
which the $N$ polymer segments are distributed with a uniform density
$\rho=N/V$. Far from the coil--globule transition and for large $N$ this
is reasonable, then the globule is expected to resemble a drop of liquid
--- the density is uniform except for a narrow interfacial region at
the surface of the globule. 
It should be borne in mind that
the assumptions behind our free energy break down at the transition
itself.
They provide an
estimate of what density of colloidal particles is required to induce
a coil--globule transition of the polymer but cannot say anything about the
critical behaviour at the transition. For a detailed study of the region
of the transition, see Ref. \cite{grassberger95} and references therein.

The starting point is a virial expansion of the Helmholtz free energy $A_g$,
of a globule, 
in the $N\rightarrow\infty$ limit.
This free energy
has had the free energy of the polymer
in the ideal coil state subtracted off.
The volume $V$ enclosed by the globule includes solvent,
which we do not treat explicitly, and colloidal spheres. These
spheres are at a density $\rho_s$ which is uniform within the globule. Then
\cite{lifshitz78,vroege92}
\begin{equation}
\frac{\beta A_g}{V}=
\rho^2B_2+\rho_s\left[\ln\rho_s-1\right]
+\rho_s^2B_2^{ss}+\rho\rho_sB_2^{ps},
\label{free1}
\end{equation}
where $\beta=1/(kT)$, for $T$ the temperature and $k$ Boltzmann's constant.
The virial coefficients $B_2$, $B_2^{ss}$ and $B_2^{ps}$ are
the second virial coefficients of the polymer--polymer,
the sphere--sphere and the polymer--sphere interactions, respectively.
We have neglected the contribution of the momentum degrees of
freedom as they do not affect the phase behaviour. It is this
which has caused the argument of the logarithm in Eq. (\ref{free1})
to have dimensions of inverse volume.
In the absence of colloid $A_g$ is just equal to the first term
on the right hand side of Eq. (\ref{free1}). This term gives the
increase in free energy with density
due to the excluded volume interactions.
The entropy cost in compressing a coil into a globule with a finite
density $\rho$ is not of order $N$ and so is not included in Eq. (\ref{free1}).
The virial coefficients are given by
\begin{eqnarray}
B_2&=&\frac{\pi}{4}P^2D\nonumber\\
B_2^{ss}&=&\frac{2\pi}{3}D_s^3\nonumber\\
B_2^{ps}&=&\frac{\pi}{4}P\left(D+D_s\right)^2.
\label{b2s}
\end{eqnarray}
$B_2^{ss}$ is the second virial coefficient of hard spheres
of diameter $D_s$. 
The above expression for $B_2^{ps}$ is the volume a rigid cylinder of
diameter $D$ excludes to a sphere of diameter $D_s$.
Equation (\ref{b2s}) for $B_2^{ps}$ therefore neglects 
the curvature of the polymer, but so long
as $P>D_s$ the polymer curves gently on a length scale of $D_s$ and
this approximation is a mild one.
The second virial coefficient of the polymer--polymer interactions
$B_2$ is obtained by splitting the polymer into segments of length
less than $P$ but much larger than $D$ and then assuming that these
interact as rigid rods \cite{vroege92}.
Consider a polymer of length $L$, we split it up into $L/l$ segments
of length $l$. The excluded volume of two cylinders of length $l$
is $(\pi/4)l^2D$ if $l\gg D$. So, the volume excluded to one segment
by the others is $(\pi/4)LlD$, and this times the number of segments
$L/l$ gives the total excluded volume, $(\pi/4)L^2D$.
Dividing this total excluded volume by $V$ and realising that
$L/P=N$ we see that we have obtained the first
term in Eq. (\ref{free1}).

The volume $V$ occupied by
the globule is within
a much larger volume of colloidal suspension which acts as a reservoir
of colloid, thus fixing its chemical potential.
The density of the colloidal spheres is different inside and outside of
a globule, thus it is more convenient to work with not the density
of the spheres but their chemical potential, which is of course always 
uniform. The coil--globule transition is then brought about by
increasing the chemical potential of the spheres.
(This is completely analogous to inducing
a coil--globule transition by reducing the temperature.)
The chemical potential $\beta\mu_s$ is equal to the derivative of the
free energy of Eq. (\ref{free1}) with respect to $\rho_s$
\begin{equation}
\beta\mu_s=\ln\rho_s+2\rho_sB_2^{ss}+\rho B_2^{ps}.
\label{mus}
\end{equation}
Under conditions of fixed $\rho$ and $\mu_s$ the correct 
thermodynamic potential is not the Helmholtz free energy but
a semigrand potential $\Omega_g$ defined by
\cite{chandler87}
\begin{equation}
\frac{\Omega_g}{V}=\frac{A_g}{V}-\rho_s\mu_s.
\end{equation}
Using Eqs. (\ref{free1}) and (\ref{mus}) this becomes
\begin{equation}
\frac{\beta\Omega_g}{V}=
\rho^2B_2-\rho_s-\rho_s^2B_2^{ss}.
\label{omega}
\end{equation}
We now go over to reduced units. 
Two reduced densities are defined:
$\zeta=\rho B_2$ and $\phi_s=\rho_s B_2/4$.
For semiflexible chains with $P\gg D$ the isotropic--nematic
transition occurs between an isotropic phase with $\zeta=5.12$ and
a nematic phase with $\zeta=5.51$ \cite{vroege92}.
$\phi_s$ is the volume fraction of colloid.
Hard spheres solidify at a volume fraction close to a half.
We have given the reduced densities at which semiflexible polymers
and spheres order as they give a good idea of the densities at which
interactions are significant.
For densities much less that those given for the transitions the
polymer/fluid of spheres is close to ideal.
Then in reduced units, Eq. (\ref{omega}) is
\begin{equation}
\frac{\beta\Omega_g}{N}=
\zeta-\frac{3}{2}\frac{\phi_s}{\zeta}\left(1+4\phi_s\right),
\label{omega2}
\end{equation}
and Eq. (\ref{mus}) is
\begin{equation}
\ln z_s=\ln\phi_s+8\phi_s+
\left(\frac{D}{P}\right)\left(1+\frac{D_s}{D}\right)\zeta,
\label{zs}
\end{equation}
where $z_s=\exp(\beta\mu_s)B_2^{ss}/4$ is a reduced
activity of the colloid.
The volume fraction $\phi_s$ of an ideal gas of colloidal
spheres in the absence of polymer is equal to $z_s$, in the
presence of excluded volume interactions $\phi_s<z_s$.
Note that Eq. (\ref{omega2}) expresses $\Omega_g$ as a function of $\rho_s$
not $z_s$. To calculate $\Omega_g$ as a function of $z_s$ we have
to solve Eq. (\ref{zs}) for $\phi_s$ at the specified value of $z_s$.
This can be done numerically.
The pressure $p$ of the globule can be obtained by taking the derivative
of $\Omega_g$, Eq. (\ref{omega2}), with respect to $\zeta$.

\section{Results and Discussion}

A mixture is specified by the values of the three length scales:
$D$, $P$ and $D_s$. 
All interactions are athermal and so the only energy scale is $kT$.
Phase behaviour is solely determined by dimensionless ratios
and the only dimensionless ratios that can be defined are then those
between lengths. The mixture's phase behaviour
is determined by two dimensionless
ratios of lengths, we choose $P/D$ and $D_s/D$. The persistence length
of DNA is around 50nm or 25 times its diameter of 2nm \cite{merchant94}.
So, we set
$P/D=25$. The ratio $D_s/D$ is set equal to 15 for our calculations.
This value is chosen as we estimate (see below) that for values of
$D_s/D$ of order 10 the continuous coil-globule transition is not
preempted by a collapse to a dense hexagonal globule.
We will return to this point when we discuss our results.

At equilibrium the pressure is uniform
which means that it must be the same inside the globule 
as in the
surrounding colloidal suspension.
A stable globule is then only possible if its pressure equals
the pressure of the surrounding suspension, which is given
by the $\zeta=0$ limit of the pressure derived from $\Omega_g$.
Local stability also requires that at that
value of $\zeta$ the pressure is an increasing function of $\zeta$.

In Fig. (2) we have plotted pressure--density plots
at two values of $z_s$, 0.5 and 1. At the smaller value of $z_s$
the pressure is a
monotonically increasing function of the density of the polymer and
so no phase with nonzero $\zeta$, i.e., no globular phase,
is stable. At $z_s=0.5$ the polymer exists as a coil.
However, for $z_s=1$ the pressure first decreases, goes through
a minimum and then increases. So, here the coil state is unstable and the 
polymer exists at a density given by the condition that its pressure
equals the pressure at $\zeta=0$.
Note that $z_s$ can easily be converted into the density
of colloid in the suspension outside the globule using the
$\zeta=0$ limit of Eq. (\ref{zs}).

As the chemical potential of the spheres is increased the slope
of the pressure versus $\zeta$ curve at $\zeta=0$ goes continuously
to zero and then becomes negative. This corresponds to a
continuous coil--globule
transition. This is seen in Fig. 3.
where we show the density of the globule as a function of $z_s$.
As the transition is approached from above, i.e., high values of $z_S$,
the density of the globule
goes continuously to zero. 
The slope of the pressure curve at $\zeta=0$ is given by the coefficient
of the term linear in $\zeta$ in the $\Omega_g$ --- the second
virial coefficient term in $\Omega_g$.
The transition is at the point when this second virial
coefficient equals 0.
There the third virial coefficient is positive and so the
$\Omega_g$ of Eq. (\ref{omega2}) has the same form as the free energy
studied by de Gennes \cite{degennes,degennes75}.

We have found a continuous coil-globule transition for our
semiflexible polymer. However, experiments on DNA
\cite{yoshikawa96,ueda96} show a first order transition;
as does a theory for polymers of long rigid segments with
short ranged attractions \cite{searb97}.
The reason for the difference is that the colloidal
spheres induce a depletion attraction with a range $\sim D_s$.
This is not much shorter than the polymer's persistence length,
in contrast to the attraction in both the DNA in the experiments
\cite{yoshikawa96,ueda96} and the model of Ref. \cite{searb97}.
When the range of the attraction is much less than the persistence
length, then the higher order virial coefficients are already
negative when the second virial coefficient becomes negative
\cite{searb97,sears97}.
The free energy is then not of the form considered by de Gennes
and others 
\cite{williams81,lifshitz78,grosberg92,cloizeux90,degennes,degennes75},
and the continuous
transition is preempted by a first order transition to
a dense globule with at least nematic ordering.
Thus, we predict that if the coil--globule transition of DNA
is induced not by altering the quality of the solvent or adding
polyvalent ions but by mixing in a colloidal suspension the
transition will be continuous not first order.

Equation (\ref{omega2}) for $\Omega_g$ is only valid when the polymer
is in the isotropic phase. We now check that the transition predicted
with this potential is not preempted by a transition to a dense
ordered phase. This is easy to do as the isotropic--nematic transition
of (pure) semiflexible chains is at a reduced pressure
$p'=\beta p B_2\simeq 26$ \cite{vroege92},
much higher than the pressure of the
transition we have found, see Fig. (2).
There is no possibility of a dense nematic or hexagonal globule
forming at this density of the colloid; the pressure inside
any such globule would be much higher than the pressure of the
colloidal suspension outside. Note that this conclusion relies on the
spheres being large, i.e., on the ratio $D_s/D$ being large.
At fixed volume fraction of the colloidal suspension, its
pressure varies as $D_s^{-3}$.
For values of $D_s$ of the order of $D$ the situation is very different.
A suspension of spheres of this size can easily be at a pressure equal
to the pressure of a dense (pure) nematic or hexagonal phase of the
polymer. In addition, the overlap of the excluded volumes of two
polymer segments only occurs when the two segments are close, within
$D_s\sim D$ of each other. By excluded volumes we mean the volumes
of space excluded by the polymer segments to the spheres \cite{vrij76}.
Thus the attraction is now short ranged, its range is much less than
$P$ and we expect a first order transition, as in Ref. \cite{searb97}.

Finally, in Fig. (4) we show the colloid density inside
and outside the globule, as a function of $z_s$.
We see that for the values of the parameters
$P/D$ and $D_s/D$ that we have taken,
a volume fraction of colloid a little more than 0.15 is
required to induce the coil--globule transition.
The colloid density outside the globule
is just that of a fluid of hard spheres at that value
of $z_s$, within the second virial
coefficient approximation. It therefore increases as $z_s$ increases.
The colloid density inside the globule, however, decreases as $z_s$ increases,
due to the increasing polymer density of the globule.

\section{Conclusion}

We have studied a system of a large, isolated semiflexible polymer molecule
in a suspension of spherical particles of diameter an order of magnitude
larger than the diameter of the polymer but much much less than the radius
of gyration of the polymer. The solvent for the polymer was good so we found
that at low densities of the spheres the polymer was a swollen coil.
However, as the density of spheres was increased beyond a certain point
the polymer underwent a coil--globule transition. The polymer molecule and
the spheres `demixed': the polymer contracted to form a dense phase
(Fig. 3) partially expelling the spheres (Fig. 4).
The driving force for the coil--globule transition is the same as that
for the demixing into two bulk phases
of long rod-like particles and spherical particles
\cite{searm97}: the excluded volume interaction between
the spheres and rods is large and this greatly reduces the
volume available to the particles. The
reduction in volume greatly reduces the
translational and rotational entropy
in phases which have high densities of both rods and spheres,
favouring demixing into two phases, each with a high density of one
component but a low density of the other.

The coil--globule transition we have found is continuous. 
Although flexible polymers such as polystyrene \cite{grosberg92,sun80}
show a continuous coil--globule transition, for DNA the transition
is discontinuous \cite{ueda96,yoshikawa96}. The DNA coil suddenly
collapses to form a dense globule with hexagonal ordering.
In the experiments which showed the discontinuous collapse of DNA,
the transition was brought about by some combination of
polyvalent ions, alcohols and small polymer molecules, see Refs.
\cite{ueda96,yoshikawa96} and references therein.
A natural assumption to make is that the collapse to a dense globule
is then brought about by a {\em short ranged} attraction, i.e.,
two DNA helices only attract each other when they are a few
nm apart. 
As discussed by the author in Refs. \cite{sears97,searb97},
the dramatic collapse to a dense ordered phase is then
not surprising, see also Refs \cite{grosberg81,doye97,bastolla97}.
The depletion attraction between polymer segments due to the presence of the
spheres is {\em long ranged},
the range is not of order $D$ but of
order $D_s$ which is an order of magnitude larger. It is this difference
in range which is changing the coil--globule transition from
discontinuous to continuous.
Thus adding colloidal particles
of size $\sim 20$nm or larger to DNA may produce a continuous collapse
of the DNA. Such a continuous collapse has not yet been seen, as far as
the author is aware.

\section*{Acknowledgments}

It is a pleasure to thank S. Fraden for bringing references
\cite{walter95,minton95} to my attention.
This work was started as a Royal Society Fellow at
the FOM institute AMOLF in Amsterdam.
I would like to thank The Royal Society for the award of a fellowship
and AMOLF for its hospitality.

%\newpage

\section*{Figure captions}
\vspace*{0.2in}

\noindent
{\bf Fig. 1}
A schematic of our semiflexible polymer (the solid curve) immersed
in a suspension of spherical particles (the shaded circles).
\vspace*{0.1in}

\noindent
{\bf Fig. 2}
Plots of the reduced pressure $p'=\beta p B_2$ 
of a globule as a function of $\zeta$, at
constant $z_s$. The solid curve (plotted on the left hand pressure
scale) is for $z_s=0.5$, and the dashed curve (plotted on the right
hand scale) is for $z_s=1$.
\vspace*{0.1in}

\noindent
{\bf Fig. 3}
The density of the globule $\zeta$ as a function of the colloid's
activity $z_s$. The density goes to 0 at the coil--globule transition,
which is at $z_s=0.515$.
\vspace*{0.1in}

\noindent
{\bf Fig. 4}
The density of the colloidal spheres $\phi_s$ outside
(solid curve) and inside (dashed curve) as a function of the colloid's
activity $z_s$.

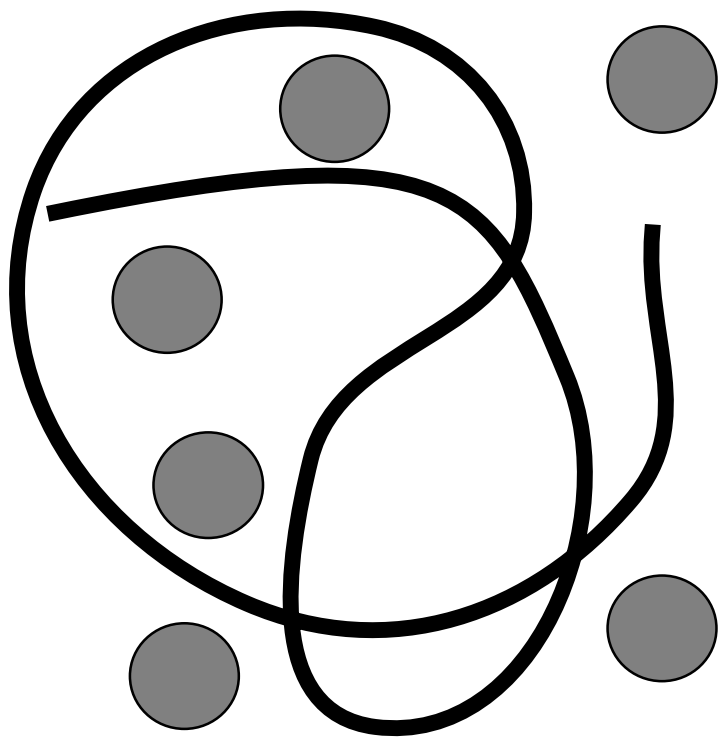
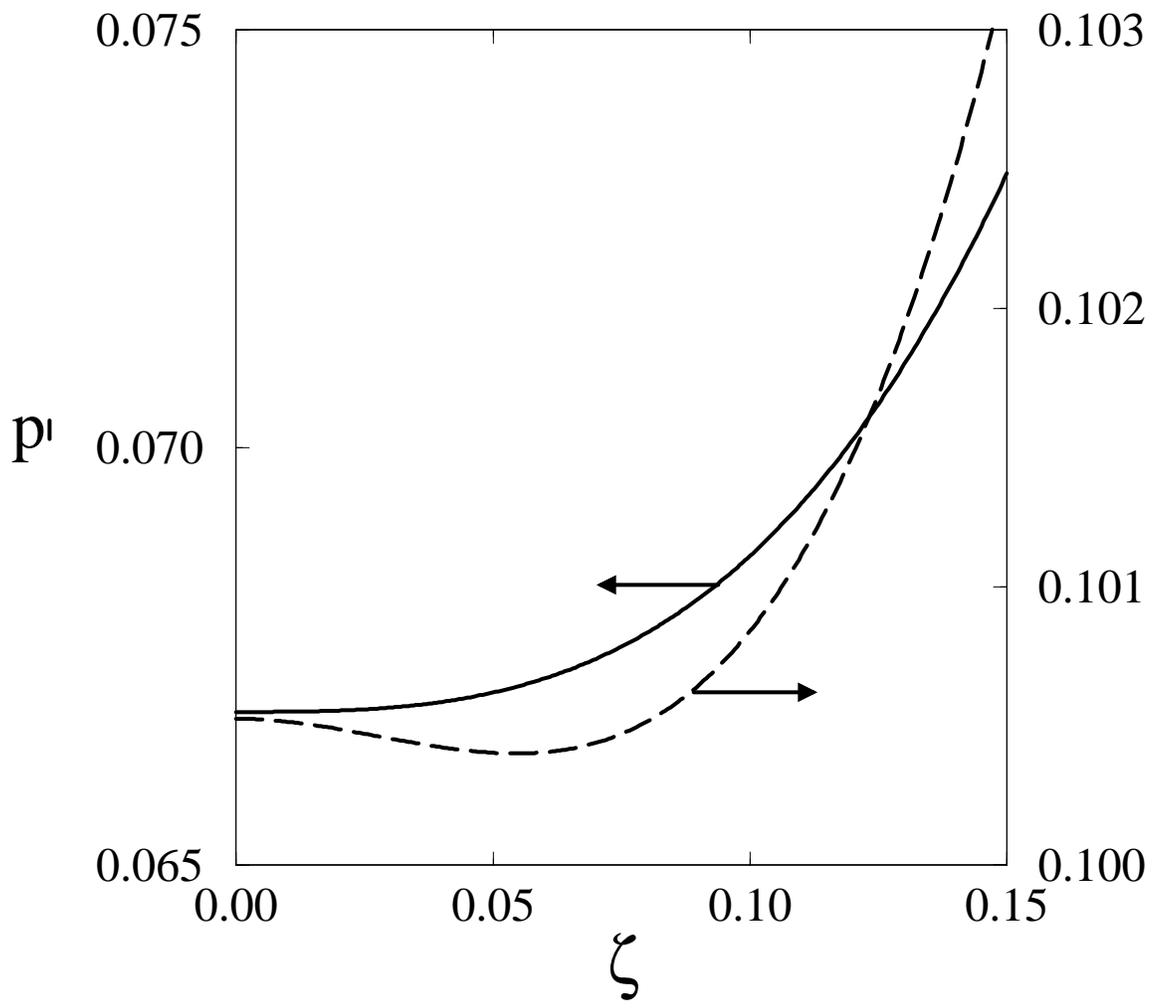
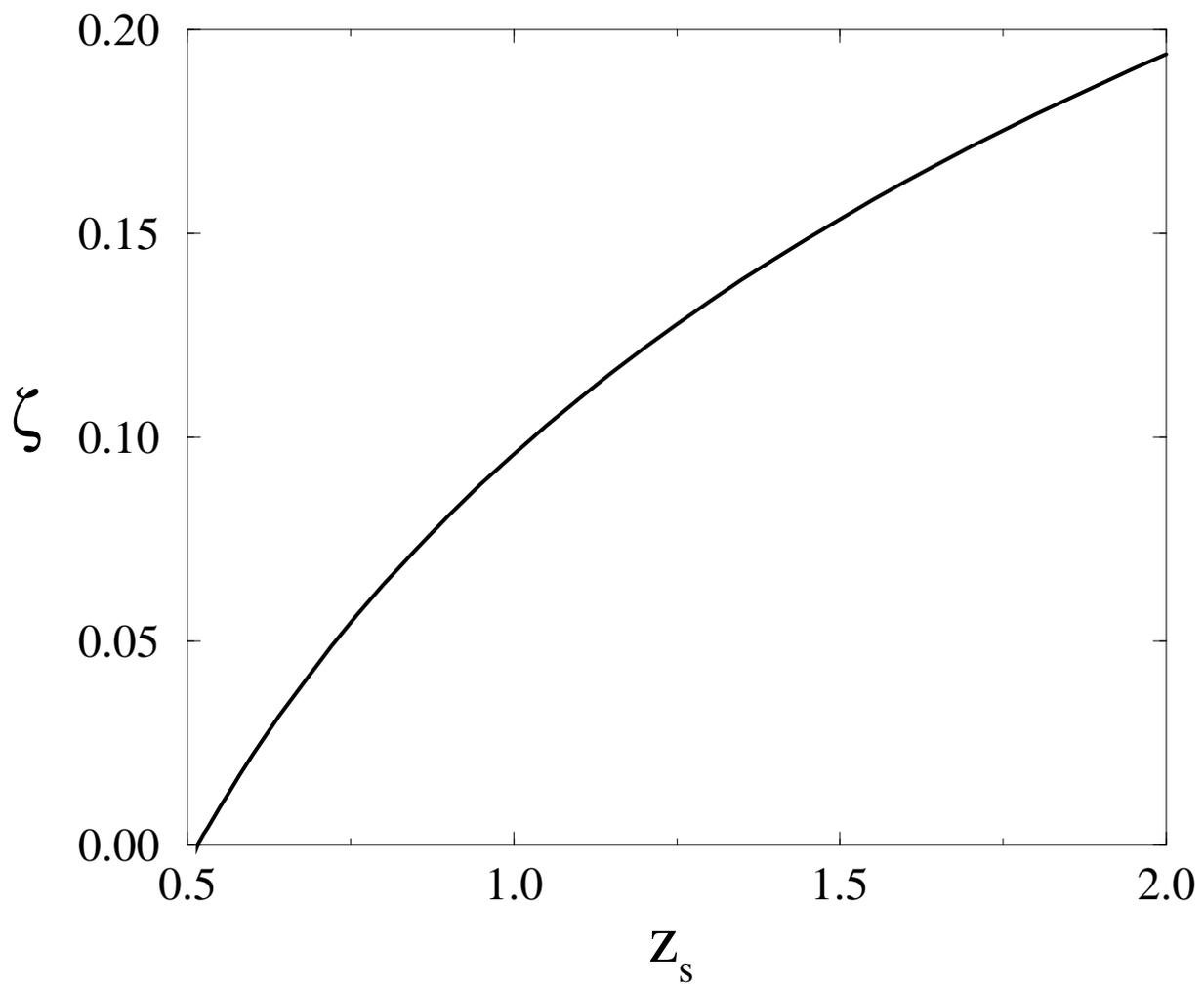
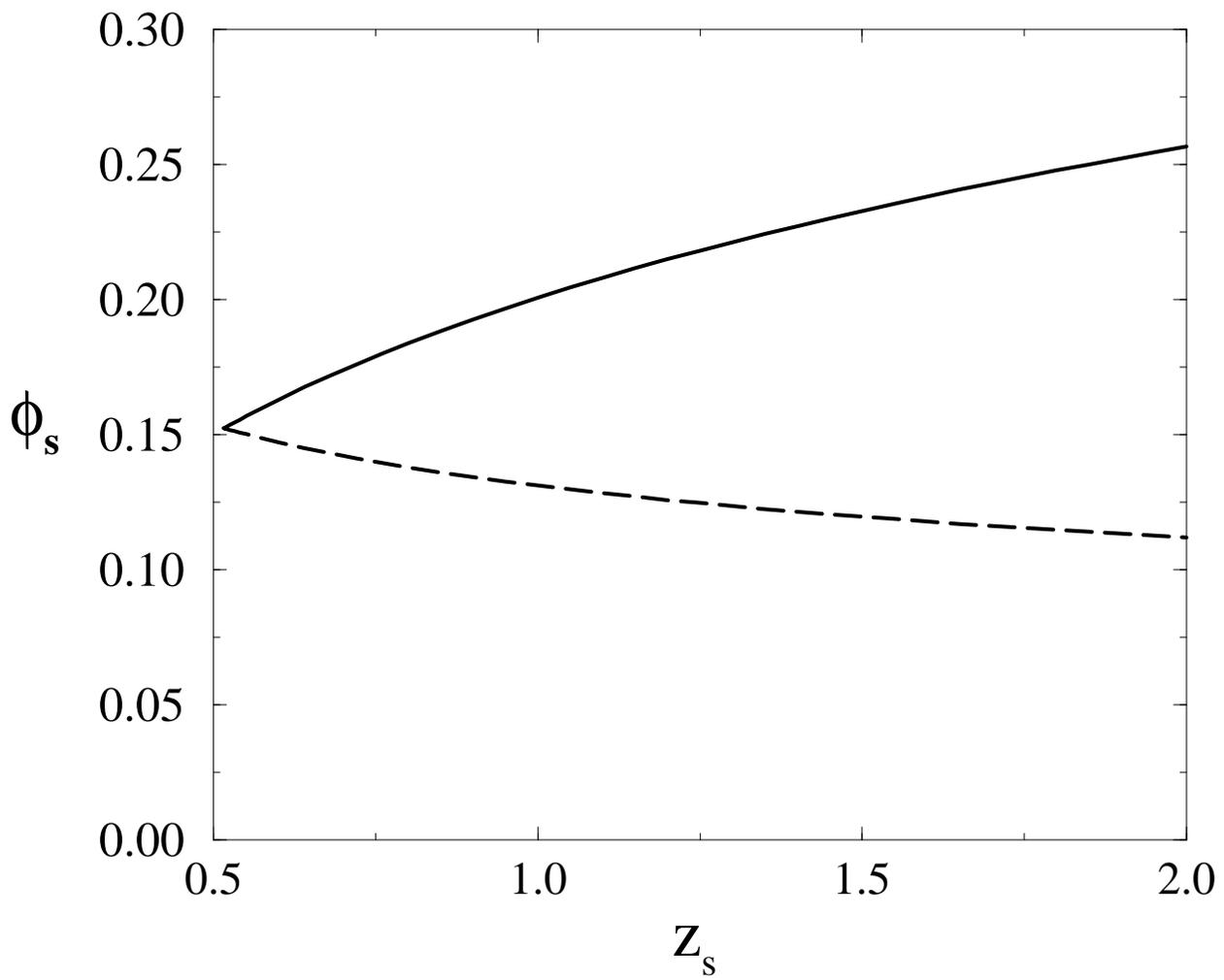
\end{document}